# Governance Challenges in Reinforcement Learning from Human Feedback: Evaluator Rationality and Reinforcement Stability

Dana Alsagheer
University of Houston
Houston , Texas, USA
dralsagh@cougarnet.uh.edu

Abdulrahman Kamal
University of California College of the Law, San Francisco
San Francisco, California, USA
Kamaaa0a@uclawsf.edu

Mohammad Kamal
Cornell Law School, Cornell University
Ithaca, New York, USA

Weidong Shi
University of Houston
Houston , Texas, USA

## Abstract

Reinforcement Learning from Human Feedback (RLHF) is central in aligning large language models (LLMs) with human values and expectations. However, the process remains susceptible to governance challenges, including evaluator bias, inconsistency, and the unreliability of feedback. This study examines how the cognitive capacity of evaluators, specifically their level of rationality, affects the stability of reinforcement signals. A controlled experiment comparing high-rationality and low-rationality participants reveals that evaluators with higher rationality scores produce significantly more consistent and expert-aligned feedback. In contrast, lower-rationality participants demonstrate considerable variability in their reinforcement decisions ($p < 0.01$). To address these challenges and improve RLHF governance, we recommend implementing evaluator pre-screening, systematic auditing of feedback consistency, and reliability-weighted reinforcement aggregation. These measures enhance the fairness, transparency, and robustness of AI alignment pipelines.

## 1 Introduction

Reinforcement Learning from Human Feedback (RLHF) has emerged as a cornerstone for aligning large language models (LLMs) with human intentions, enabling adaptive behavior beyond hardcoded objectives. Prominent models such as GPT-4, Claude, Bard, and LLaMA 2-Chat rely heavily on RLHF to refine their outputs through human preference signals [2, 15]. The RLHF pipeline typically involves three core stages: collecting human feedback, training a reward model to predict that feedback, and optimizing the model policy via reinforcement [4, 18] (Figure 1).

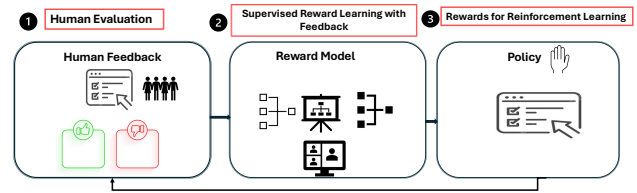

**Figure 1: A Structured Framework for Reinforcement Learning: Integrating Human Feedback, Reward Modeling, and Policy Optimization.**

Despite its widespread adoption, RLHF is not without risk. As AI systems take on increasingly high-stakes responsibilities—ranging from legal reasoning to content moderation—the reliability of human evaluators becomes a critical point of failure. Human feedback is often inconsistent, cognitively biased, or misaligned with expert judgment [7, 12]. These weaknesses are exacerbated when feedback originates from individuals with limited reasoning capacity or cultural homogeneity, leading to volatile and potentially adversarial reinforcement signals.

Current RLHF pipelines rarely include governance safeguards to evaluate the quality of human input. Without robust auditing and evaluator vetting, models trained on unfiltered human feedback may reflect irrational, biased, or unstable behavior, undermining trust and generalizability. As LLMs' capabilities continue to scale, governance strategies must shift from merely collecting feedback to actively managing their quality and representativeness.

This study addresses a critical gap in RLHF governance: the role of human rationality in shaping the stability and fairness of reinforcement signals. We present empirical evidence showing that high-rationality evaluators generate significantly more consistent and expert-aligned feedback compared to their lower-rationality counterparts. These findings raise urgent questions about who should serve as evaluators in alignment pipelines and what mechanisms are needed to ensure trustworthy RLHF.

Our contributions are threefold:

(1) We quantify the impact of evaluator rationality on the consistency of reinforcement signals, using real-world rationality tasks and expert-labeled benchmarks.
(2) We identify key governance risks arising from unqualified or demographically homogeneous evaluators and propose mechanisms to mitigate these risks.
(3) We introduce a governance framework for RLHF that includes evaluator pre-screening, feedback consistency auditing, and reliability-weighted aggregation.

These interventions offer a path forward for improving AI alignment systems' fairness, transparency, and robustness. This work contributes to a growing body of literature calling for more human-centered and governance-aware approaches to designing reinforcement learning pipelines.

## 2 Related Work

Reinforcement Learning from Human Feedback (RLHF) is widely used to align large language models (LLMs) with human preferences. However, research highlights its limitations, including hallucinations[10, 26], biased model responses[16, 21], and sycophantic behavior—where models optimize for agreement rather than correctness [17]. RLHF also poses privacy risks, as models may memorize and leak sensitive data [5, 11]. Furthermore, it has failed to prevent adversarial attacks, such as jailbreaking and prompt injection, which threaten real-world security [1, 13, 23, 24]. Alternative approaches have been proposed to address these challenges. Constitutional AI[2] integrates predefined principles to improve alignment, while adversarial training[8] strengthens model robustness against manipulation. Other methods, such as human-in-the-loop evaluation[6] and multi-step reward modeling[22], seek to enhance reliability. However, these solutions do not fully resolve RLHF's limitations, as they still depend on human feedback, which is prone to biases, inconsistencies, and rationality gaps.Building on prior research, this work systematically assesses RLHF's governance failures, focusing on evaluator reliability, transparency, and fairness. Unlike previous studies that focus on technical refinements, we examine the structural deficiencies of RLHF, highlighting the risks of low-rationality evaluators and proposing governance mechanisms to improve reinforcement consistency and bias mitigation.

## 3 Methodology

To examine how the selection of human evaluators influences the consistency and objectivity of reinforcement learning signals, we conducted a two-stage online experiment with ten participants, each holding at least a bachelor's or master's degree. The goal was to assess how reliably humans evaluate model-generated outputs and to quantify potential biases in their feedback.

### 3.1 Participant Grouping and Rationality Assessment

Participants first completed a 20-item rationality test adapted from Burgoyne et al. [3] to evaluate cognitive reflection and reasoning ability. Based on test performance, participants were stratified into groups representing varying levels of rational reasoning expertise. Higher scores were used as a proxy for greater evaluative competence.

### 3.2 AI Response Evaluation Task

Each participant then evaluated 25 AI-generated responses from GPT-4 [14] on a new set of multiple-choice rationality questions. Participants assessed each AI answer for correctness, even when the response differed from their judgment. A separate set of 25 questions was also generated by GPT-4 using the OpenAI API with default parameters (e.g., temperature, top-p) to ensure unbiased generation conditions. Participants evaluated these AI-generated questions to assess the robustness of their reinforcement signals under less familiar or less structured conditions.

### 3.3 Metrics for Consistency and Bias

To quantify the quality of human feedback, we introduced two core metrics: Test-Retest Consistency Score (TRCS) and Bias Deviation (BD).

**Test-Retest Consistency Score (TRCS)** measures the internal stability of each evaluator's feedback across two evaluation rounds on the same set of model outputs:

$$TRCS = \frac{\text{Number of Unchanged Responses}}{\text{Total Responses}}, \quad (1)$$

where higher values indicate greater consistency and decision stability over time.

**Bias Deviation (BD)** captures the extent to which individual evaluators deviate from expert-aligned ground truth. It is computed as the average absolute difference between the evaluator's binary reinforcement signal ($F_i$) and the expert-annotated ground truth label ($G_i$), across $N$ questions:

$$BD = \frac{1}{N} \sum_{i=1}^{N} |F_i - G_i|. \quad (2)$$

A BD score of 0 indicates perfect alignment with expert judgment, while higher values reflect increasing divergence and potential evaluator bias.

A psychology Ph.D. student with domain expertise in rationality assessment created all expert labels independently. To ensure transparency and reproducibility, the full dataset of questions—both adapted and AI-generated—will be made available in our GitHub repository.

## 4 Results and Analysis

### 4.1 Consistency in Reinforcement Signals

Participants who performed well on pre-screening tests exhibited significantly higher feedback stability, with a 92%

Table 1 presents the TRCS results for both groups.

Table 1: Test-Retest Consistency Score (TRCS) Across Evaluator Groups

| Group | TRCS Mean | Standard Deviation |
| --- | --- | --- |
| High-Rationality | 0.92 | 0.05 |
| Low-Rationality | 0.45 | 0.17 |

## 4.2 Bias Deviation in Reinforcement Decisions

The results in Table 2 show that high-rationality evaluators exhibited significantly lower bias deviation (BD = 0.08) compared to the low-rationality group (BD = 0.34), with a notable difference in standard deviation. This indicates that legal experts provided more consistent and reliable reinforcement signals, while general population participants demonstrated more significant variability, leading to potential biases in RLHF.

Table 2: Bias Deviation in Reinforcement Decisions

| Group | Bias Deviation (BD) | Standard Deviation |
| --- | --- | --- |
| High-Rationality | 0.08 | 0.04 |
| Low-Rationality | 0.34 | 0.12 |

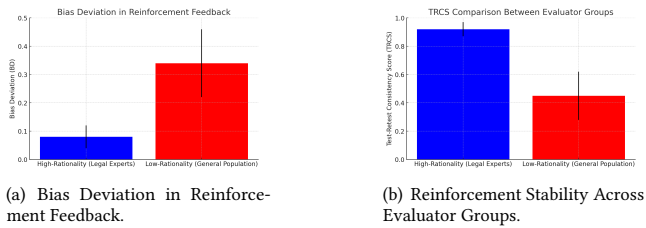

(a) Bias Deviation in Reinforcement Feedback.

(b) Reinforcement Stability Across Evaluator Groups.

Figure 2: Comparison of bias deviation and reinforcement stability across evaluator groups.

## 5 Discussion

Our results demonstrate that evaluators' selection and cognitive aptitude are critical in the quality and consistency of reinforcement learning from human feedback (RLHF). Specifically, we observed that participants with higher rationality scores delivered significantly more stable and expert-aligned feedback, as reflected in both the Test-Retest Consistency Score (TRCS) and the lower Bias Deviation (BD) from expert ground truth. This suggests that rational reasoning ability is a desirable trait and a foundational requirement for evaluators in alignment pipelines.

These findings highlight a broader concern in RLHF design: not all human feedback is equal. The RLHF process risks introducing systematic biases without rigorous pre-screening or qualification metrics. This is particularly problematic when evaluators are selected for economic efficiency rather than evaluative competence. Outsourcing RLHF tasks to lower-cost labor markets—such as the Philippines or Kenya—is now standard industry practice. Yet, it creates a monoculture of feedback that reflects limited cultural, linguistic, and epistemic perspectives [19, 20]. Such demographic homogeneity may distort the development of "aligned" AI by overfitting models to the dominant worldview of a narrow population of annotators.

Moreover, our findings challenge the assumption that large numbers of annotators inherently yield better feedback. When low-rationality evaluators are included, aggregation may dilute the expert signal and amplify noise, particularly if simple majority voting mechanisms are used. Reinforcement feedback must therefore be weighted by evaluator reliability, not treated uniformly. This reinforces the need to integrate competence-based pre-screening and post-hoc consistency auditing into RLHF pipelines.

A further implication concerns the interpretability of the bias metric itself. We proposed that the Bias Deviation (BD) score quantifies how far an evaluator's feedback diverges from expert-labeled responses. Importantly, this deviation is not treated as mere disagreement but as a signal of potential misalignment. While multiple valid perspectives exist in subjective domains, in structured rationality tasks, such as those used here, there exists a clear normative ground truth. Therefore, the BD metric is a proxy for epistemic alignment, not just preference diversity. This is crucial for RLHF in domains like law, medicine, and public policy, where correctness is not purely subjective.

Future governance frameworks must embed these evaluative safeguards more deeply to build more representative and fair AI systems. Technical fixes—such as adversarial input filtering, outlier suppression, and ensembling—can reduce some inconsistencies, but they are no substitute for human-centered design. As we argue, integrating HCI-informed solutions such as reputation tracking, diversified feedback sourcing, and feedback quality visualization is essential to ensure the trustworthiness of RLHF processes.

We further advocate for a transformative shift toward decentralized evaluator selection. Integrating Decentralized Autonomous Organizations (DAOs) with blockchain-backed audit trails offers a promising direction. DAOs can support the transparent recruitment, ranking, and compensation of evaluators based on their historical reliability and specialization. Smart contracts can manage evaluator incentives, ensure dispute resolution, and automate the removal of consistently biased actors. This structure empowers contributors globally, moving beyond geographic bias toward a more skill- and value-aligned feedback ecosystem [9, 25].

In essence, we propose a future where AI alignment is governed not by opaque corporate hierarchies, but by transparent, decentralized, and meritocratic systems that value high-quality human judgment. Our findings suggest that such reform is possible—and necessary—for the next generation of trustworthy AI.